\documentstyle[prl,aps,multicol,epsf]{revtex}
\begin{document}
\newcommand{\e}[2]{\begin{equation}#2 \label{#1} \end{equation}}
\newcommand{\ea}[4]{\begin{eqnarray} \label{#1} #2\\ \label{#3} #4 
                    \end{eqnarray}}
\newcommand{\te}[1]{\mbox{$\, #1 \,$}}
\newcommand{\et}{\mbox{$\; = \;$}}
\newcommand{\plu}{\mbox{$\; + \;$}}
\newcommand{\mi}{\mbox{$\; - \;$}}
\newcommand{\Av}{\mbox{$\mathbf{ A}$}}
\newcommand{\ps}{\mbox{$\psi (\mathbf{ r})$} }
\newcommand{\Z}{\mbox{$Z\!\!\! Z$}}
\newcommand{\Nn}{\mbox{$I\!\!N_0$}}
\newcommand{\bi}[3]{\begin{figure}
                    \begin{center}
                    \epsfxsize=#2
                    \begin{minipage}{\epsfxsize}
                    \epsffile{#1.ps}
                    \label{#1}
                    \end{minipage}
                    \begin{minipage}{\linewidth}
                    \setlength{\parindent}{0cm}
                    \caption{\it #3}
                    \end{minipage}              
                    \end{center}
                    \end{figure}  }
\draft 
\title{Critical fields  of mesoscopic superconductors}
\author{ Robert Benoist and Wilhelm Zwerger}
 \address{Ludwig-Maximilians-Universit\"at, Sektion Physik,
Theresienstra\ss e 37, D-80333 M\"unchen, Germany}
\date{\today}
\maketitle
\begin{abstract}
Recent measurements have shown oscillations in the upper critical
field of simply connected mesoscopic superconductors. A quantitative
theory of these effects is given here on the basis of a
Ginzburg-Landau description. For small fields, the $H-T$ phase
boundary exhibits a cusp where the screening currents change sign for
the first time thus defining a lower critical field $H_{c1}$. In the limit where many flux
quanta are threading  the sample, nucleation occurs at the boundary
and the upper critical field becomes identical with the surface
critical field $H_{c3}$.
\end{abstract}
\vskip0.2cm

\begin{multicols}{2}
\section{Introduction}
\footnotetext{Dedicated to Prof.~W.~G\"otze on the occasion of his
60th birthday.} As is well known, superconductivity only exists
at sufficiently low temperatures $T$ and small external magnetic
fields $H$. The resulting $H-T$ boundary for the normal to superconducting transition is
determined by the Ginzburg parameter \te{\kappa=\lambda/\xi}. For type II 
superconductors with \te{\kappa>1/\sqrt{2}} one obtains a lower
(\te{H_{c1}}) and an  
upper (\te{H_{c2}}) critical field which - for bulk samples - are universal 
functions of temperature \cite{deGennes,Tinkham}. However, it was
realized long ago by Saint-James and de Gennes
\cite{SolGennes,SaintJamesDeGennes} that in the presence of a surface,
these results 
are changed considerably. Regarding \te{H_{c1}}, there is a surface barrier for
the entrance of the first flux quantum. Thus the field up to which
the sample stays in the Meissner phase may be much larger than the 
thermodynamic \te{H_{c1}} \cite{SolGennes}.  In the case of the upper
critical field, 
superconductivity in a bounded sample persists even in the range
\te{H_{c2}<H<H_{c3}=1.69H_{c2}} provided the external field is parallel to the
surface \cite{SaintJamesDeGennes}. In this regime only a thin sheet at
the sample boundary of the order of the zero field coherence length
\te{\xi}  is superconducting. Quite generally, in 
samples whose size is of the order of the $T=0$ coherence length, one
expects that the $H-T$ phase boundary will strongly depend on the
detailed form of the sample, reflecting the possible eigenmodes for
the complex superconducting order parameter $\psi (\bf{{\bf r}})$ in the
given geometry. Experimentally this was recently studied by
Moshchalkov {\it et al.}~\cite{Moshchalkov}, who investigated the
temperature dependence of
the upper critical field of small mesoscopic aluminium samples
with typical sizes of 1$\mu$ or less. The observed $H-T$ phase boundary
turned out to exhibit very peculiar size effects. Specifically it was
found that Aharonov-Bohm like oscillations in the critical field were
present even in a simply connected geometry, similar to the
Little-Parks oscillations found long ago \cite{LittleParks} in thin
walled superconducting cylinders.

It is the purpose of this work to investigate size effects in the
critical fields of small superconductors on the basis of a
Ginzburg-Landau (GL) theory. We will find that such a description
apparently remains valid down to system sizes of only a few
coherence lengths. By a careful solution of the boundary value problem
for the linearized GL-equation near $T_c$, we are able to quantitatively
describe the observed structure in the upper critical field of a small
disc. In the limit where many flux quanta are threading
the sample, the upper critical field is in fact a surface critical field,
reproducing the standard \te{H_{c3}} value of a semi-infinite
geometry. Moreover we determine a generalized lower critical field for
mesoscopic discs and rings.
An interesting point is that 
the
eigenvalue spectrum which determines the suppression of the critical
temperature $T_c (H)$ as a  function of magnetic field, is rather
different from the case of electron levels in a quantum dot, because
of the different boundary conditions.
\section{Critical fields of superconducting discs and rings}
Let us consider a small disc with radius $R$ and thickness $d$ in an
external magnetic field \te{{\bf H}=H{\bf e}_z}, which is
perpendicular to the sample surface at \te{z=\pm d/2}. Near the normal
to superconducting transition the change in the free energy with
respect to the normal state can be expressed in terms of a
GL-functional of the complex superconducting order parameter \te{\psi
({\bf r})} \cite{deGennes,Tinkham} 
\begin{eqnarray}
\label{frengl}
F[\psi]  \et & F_{n} + \int \limits_V 
\left \{ \frac{\hbar^2}{4\mu} \left | \left (\nabla \mi
\frac{2ie}{\hbar c}\Av \right )? \psi({\bf r}) \right |^2 \right .\\
&\left. \plu  a|\psi({\bf r})|^2 \plu  \frac{b}{2}|\psi({\bf r})|^4 \plu 
\frac{{\bf B}^2}{8\pi} \right \}  \,d^3r \; . 
\nonumber\end{eqnarray}
Here \te{{\bf B}=\nabla \times \Av} is the  magnetic field in the
sample with volume $V$, \te{a=a' (T-T_c)/T_c} and $b$ are the standard
GL-coefficients and $\mu$ the effective electron mass \cite{deGennes,Tinkham}.
In principle there are also surface contributions to
the free energy functional (\ref{frengl}) which may be important for mesoscopic
samples with a large surface to volume ratio. In our treatment below
such contributions are neglected, which is justified only a posteriori.
In the vicinity of the transition the order parameter and the
screening currents are small. To lowest order we may therefore neglect
the quartic term in $F[\psi]$ and replace the magnetic field by the
external one. The most probable configuration of the order parameter
which follows from the mean field equation \te{\delta F[\psi]/\delta
\psi^* =0} is then determined by the eigenvalue problem
\begin{equation}
- \frac{\hbar^2}{4\mu} \left ( \nabla - \frac{2ie}{\hbar c}\Av \right
)^2 \psi \et \mi  a \psi 
\label{glg}
\label{glgav}
\end{equation}
for a particle with charge $2e$ in an external magnetic field
(\te{e<0}). Assuming that the sample is embedded in an insulating
medium, the relevant boundary condition is that of vanishing current
normal to the sample surface \te{\partial V}. In covariant form the
corresponding  Neumann boundary condition is \cite{deGennes}
\begin{equation}
\left . {\bf n} \cdot \left (\nabla - \frac{2\, i \, e}{\hbar \, c} \Av \right ) \psi \right \arrowvert_{\partial V} = \; 0 ,
\label{randbed}
 \end{equation}
where \te{\bf n} is a unit vector normal to the sample surface. In
order to determine the $H-T$ phase boundary, we must find the lowest
eigenvalue \te{E_0 (H)} associated with a nonzero order parameter \te{\ps
\neq 0}. From \te{E_0 (H)} the transition from the normal to the
superconducting state is determined by 
\e{aeh}{-a= a' \frac{T_c - T_c (H)}{T_c}= E_0 (H).}
Here \te{T_c} is the (mean field) transition temperature of the
infinite system with zero field. Since \te{E_0 (H) \geq E_0 (0)} quite
generally \cite{Simon}, the transition temperature at finite field is
always smaller or equal than at \te{H=0}. In order to treat the case
of discs or rings, it is convenient to introduce cylindrical
coordinates \te{(\rho, \phi, z)}. In the appropriate gauge \te{\Av = 
 H \rho \,{\bf e}_{\phi}/2} the solution of the Schr\"odinger  equation
(\ref{glg}) can then be written as 
\e{ansatzpsisymm}{\psi = {\cal R} (\rho) \;  e^{i m \phi} \;
e^{i  k_{\nu}  z} .}
Here \te{m \in \Z} is the angular momentum quantum number and
\te{k_{\nu}= \nu \pi/d} with \te{\nu \in \Nn} the discrete wavevector for 
motion in the $z$-direction. Since the lowest eigenvalue has always
\te{\nu=0}, we will omit the $z$-dependence and the associated quantum
number $\nu$ in the following. Introducing a dimensionless variable
\te{\zeta= \rho^2/2l^2_H} with \te{l_H= (\hbar c / 2 |e| H)^{1/2}}
the magnetic length for charge \te{2e}, the differential equation for
\te{{\cal R}(\zeta)} can be reduced to Kummer's confluent
hypergeometric equation \cite{Abramowitz}
\e{dglxi}{\zeta \frac{\partial^2 w}{\partial \zeta^2} \plu \left(|m|+1
\mi \zeta \right ) \frac{\partial w}{\partial \zeta}
\mi  \alpha \; w \; \et \; 0 }
by the substitution \te{{\cal R} (\zeta ) \et e^{- \frac{\zeta}{2}} \; \zeta^{\frac{|m|}{2}}
\; w (\zeta)}. The dimensionless parameter $\alpha$ is  directly
related to the eigenvalue \te{E(H)} by
\e{alvh}{\alpha = - \frac{E(H)}{\hbar \omega_c} \plu \frac{1}{2} (|m|+m+1)}
with \te{\omega_c= |e|H/\mu c} the standard cyclotron frequency. Using
\te{-E_0 (H)=a} and \te{a'=\hbar^2/4\mu \xi^2(0)} with \te{\xi (0)}
the zero temperature GL-coherence length, the maximum value of $\alpha$ --
which always has  \te{m \leq 0} -- determines the magnetic
field shift of the transition temperature by
\e{dt}{\frac{T_c - T_c (H)}{T_c} = \left [ \frac{1}{2} - \alpha_{max} (H)
\right ]\frac{4 \tilde{\Phi}}{\Phi_0}}
with \te{\tilde{\Phi}= \pi \xi^2(0)H}  and \te{\Phi_0= hc/2|e|} the superconducting flux quantum.
In an infinite sample the ground state is the lowest Landau level with
\te{E_0^{\infty}= \hbar \omega_c /2}, i.e.~\te{\alpha^{\infty}=0}. The
phase boundary is then given by 
\e{dtbulk}{\frac{T_c - T_c (H)}{T_c}= 2
\frac{\tilde{\Phi}}{\Phi_0},} 
which is equivalent to the standard relation \te{H=H_{c2} (T)=
\Phi_0/2 \pi \xi^2 (T)} \cite{deGennes,Tinkham}. For the finite
system, the spectrum of eigenvalues follows from the Neumann boundary
condition (\ref{randbed}) at the inner \te{(R_i)} and outer
\te{(R)} radius of the ring. The general solution of (\ref{dglxi})
is a linear combination of Kummer functions \cite{Abramowitz}. In the
case of a disc geometry only
\e{soldisc}{w_1 (\zeta) = \Phi (\alpha,|m|+1,\zeta)= _1F_1(\alpha,|m|+1,\zeta)}
is allowed since the second linear independent solution diverges at
the origin. Using standard recursion relations for \te{_1F_1}, it is
straightforward to show that the boundary condition at \te{\rho=R},
which simply reads \te{dR/d\rho=0}, since \te{\Av \cdot {\bf n}=0},
leads to  
\ea{randbednum}{\nonumber(|m|+1-\alpha)\, \Phi({\alpha-1},{|m|+1},{\zeta_R})&
}{randbedz}{\mi 
\Phi ({\alpha},{|m|+1},{\zeta_R}) \plu  \alpha \,
\Phi({\alpha+1},{|m|+1},{\zeta_R})&\, = 0 } 
with \te{\zeta_R=R^2/2l^2_H}. For each given \te{m} equation
(\ref{randbednum}) determines a discrete series of eigenvalues 
\te{\alpha_{n m} (H)}, \te{n\in \Nn}, which are \underline{de}creasing with increasing
\te{n}. They obey \te{\alpha_{nm}\leq 1/2} and are continuous in \te{\zeta_R= \Phi / \Phi_0}
\cite{Benoist} which is just the external flux \te{\Phi= \pi R^2 H}
through the area of the disc in units of the flux quantum.
Analytical results for the spectrum can be obtained in the low field
limit \te{\Phi \rightarrow 0}. In this limit it is straightforward to
treat the general case of a ring with \te{\sigma= R_i/R \leq 1}.
Standard second order perturbation theory in the magnetic field then
leads to a shift in the transition temperature which is given by
\e{dtzero}{\frac{T_c - T_c (H)}{T_c}= \frac{1}{2} (1 + \sigma^2)
\frac{\tilde{\Phi} \Phi}{\Phi_0^2} + ...\; .}
The corrections to this result are of order  \te{\Phi^4}, since the
ground state energy is even in \te{\Phi}.
For a very thin ring with \te{\sigma \rightarrow 1^-} this agrees with
the low field limit of the Little-Parks result \cite{LittleParks}
\e{lipa}{\frac{T_c - T_c (H)}{T_c}= \frac{\xi^2(0)}{R^2} \min_{m \in
\Z} \left | m - \frac{\Phi}{\Phi_0} \right |^2,}
as expected. For general magnetic fields the phase boundary can only
be obtained numerically. To this end we have directly solved the
transcendental equation (\ref{randbednum}) which allows us to
determine the spectrum without any discretization error.
The energy levels are thus obtained with arbitrary accuracy, in
contrast to previous work by Saint-James \cite{SaintJames} or by
Nakamura and Thomas \cite{Nakamura} who consider Dirichlet boundary
conditions. The results are shown in Fig.~1, where the dimensionless
eigenvalues \te{\frac{4 \mu R^2}{\hbar^2} E_{n m}} for \te{n=0} and
\te{m=2,1,0,-1,...,-10} are plotted as functions of \te{\Phi /\Phi_0}.
\bi{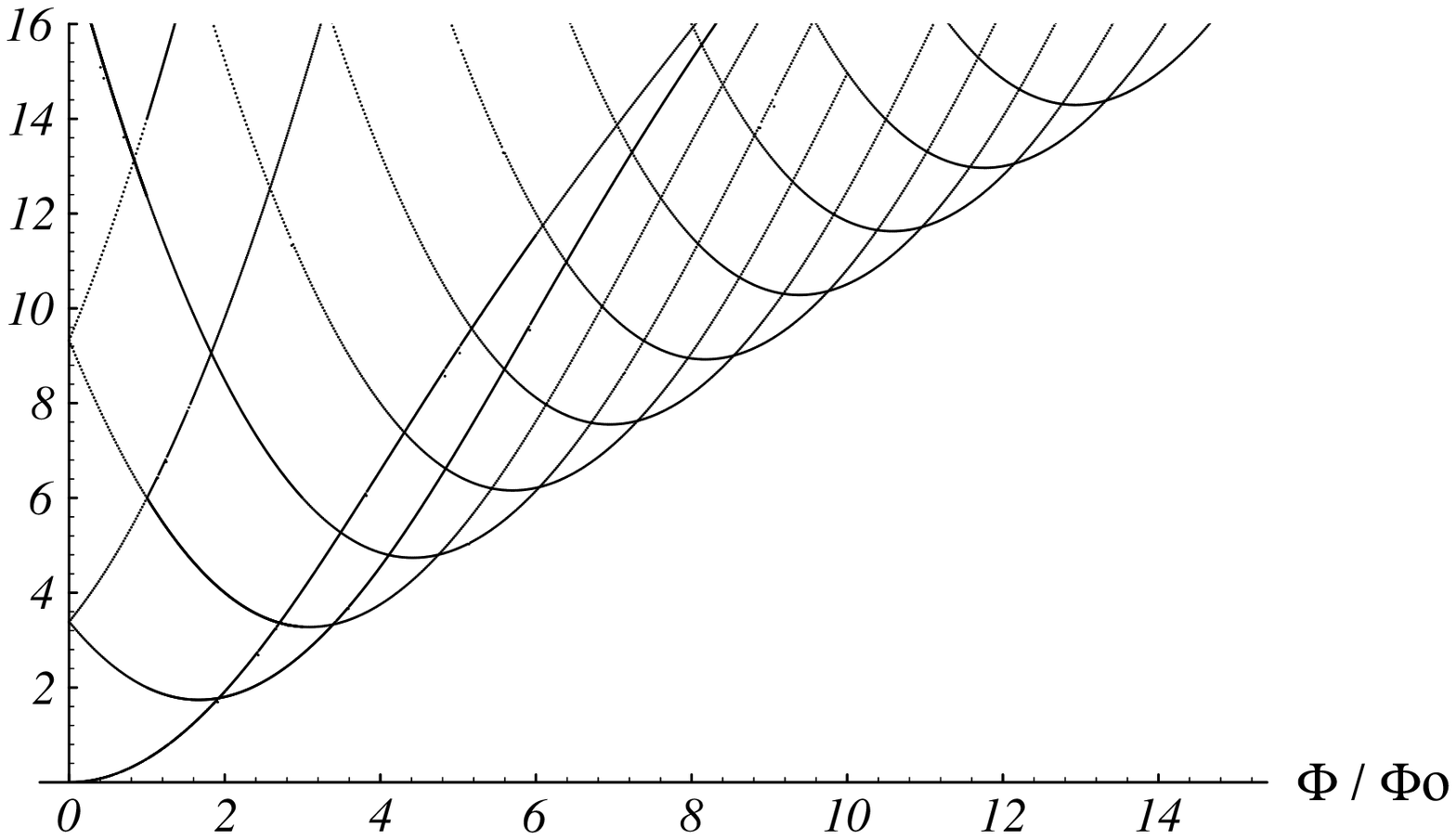}{1.1\linewidth}{Dimensionless energy eigenvalues of a disc as a function of
the external magnetic flux.}
Evidently the lowest eigenvalue exhibits an oscillatory behaviour with
cusps at values \te{\Phi^{(j)}, \, j=1,2,...} where the magnetic
quantum number of the lowest eigenstate jumps by one unit.
The dimensionless distances between successive cusps 
\e{defdelta}{\Delta_j= \frac{\Phi^{(j)} \mi \Phi^{(j-1)}}{\Phi_0}\qquad
(\Phi^{(0)}=0)} 
are given in table 1 with an accuracy corresponding to the last given
digit.

\vspace{0.2cm}

\begin{tabular}{|llc|cll|}
\hline
$\Phi^{(1)}\phantom{1}=$ & $\phantom{1}1.923765$&$\Phi_0$ & $\qquad$ & $\Delta_1=$ &
$1.923765$  \\ \hline
$\Phi^{(2)}\phantom{1}=$ & $\phantom{1}3.392344$&$\Phi_0$ & $\qquad$ & $\Delta_2=$ &
$1.468579$  \\ \hline
$\Phi^{(3)}\phantom{1}=$ & $\phantom{1}4.747920$&$\Phi_0$ & $\qquad$ & $\Delta_3=$ &
$1.355676$  \\ \hline
$\Phi^{(4)}\phantom{1}=$ & $\phantom{1}6.045882$&$\Phi_0$ & $\qquad$ & $\Delta_4=$ &
$1.297962$  \\ \hline
$\Phi^{(5)}\phantom{1}=$ & $\phantom{1}7.3068$&$\Phi_0$ &  $\qquad$ & $\Delta_5=$ &
$1.2609$   \\ \hline
$\Phi^{(6)}\phantom{1}=$ & $\phantom{1}8.5423$&$\Phi_0$ &  $\qquad$ & $\Delta_6=$ &
$1.2355$   \\ \hline
$\Phi^{(7)}\phantom{1}=$ & $\phantom{1}9.7584$&$\Phi_0$ &  $\qquad$ & $\Delta_7=$ &
$1.2161$   \\ \hline
$\Phi^{(8)}\phantom{1}=$ & $10.9591$&$\Phi_0$ &  $\qquad$ & $\Delta_8=$ &
$1.2007$   \\ \hline
$\Phi^{(9)}\phantom{1}=$ & $12.1477$&$\Phi_0$ &  $\qquad$ & $\Delta_9=$ &
$1.1886$   \\ \hline
$\Phi^{(10)}=$ & $13.3255$&$\Phi_0$ &  $\qquad$ & $\Delta_{10}=$ &
$1.1778$   \\ \hline
\end{tabular}

TABLE 1

\vspace{0.2cm}

Experimentally the oscillatory behaviour of the ground state energy
is directly reflected in the $H-T$ phase boundary.
For the case of a disc discussed here, this was actually first
observed by Buisson {\it et al.}~\cite{Buisson}. In their experiment,
however, the presence of two gold contacts led to a boundary condition
which is different from (\ref{randbed}) over part of the sample
boundary. While the oscillations were still present, a detailed
comparison with theory was difficult (for instance the
first cusp was observed at \te{\Phi\approx 2.5 \Phi_0} compared
to \te{\Phi = 1.924 \Phi_0} in the pure Neumann case). The more
recent experiments of Moshchalkov {\it et al.}~\cite{Moshchalkov}, however,
are in very good agreement with the theoretical predictions.
This may be seen from a comparison with the measured deviation of the
temperature shift \te{\Delta T_c=T_c \mi T_c (H)} from the average
linear behaviour which is shown in Fig.~2. 
\bi{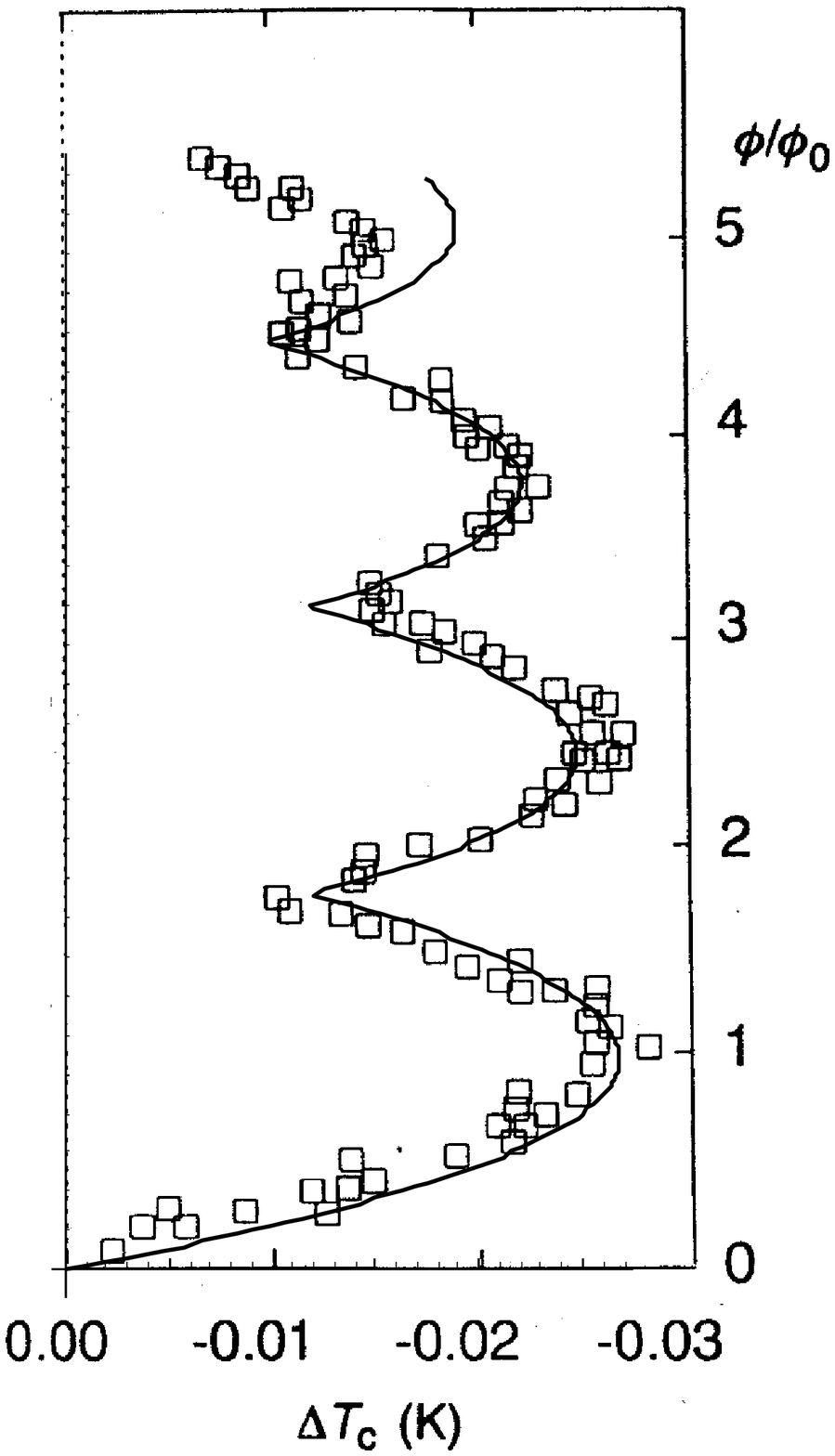}{\linewidth}{Flux dependence of the oscillatory part of
the temperature shift. The experimental data (squares) is  taken from
\protect\cite{Moshchalkov}, the solid line 
is the theoretical prediction based on equation (\protect\ref{dt}).}
Here we have used the
experimental value \te{\xi(0)=1 \mu}, and a disc area which is only 2.7 \%
smaller than the area of the almost rectangular sample used in the
experiment. It is important to note that the periods \te{\Delta_j}
decrease monotonically from \te{\Delta_1=1.924} to \te{\Delta_{\infty}=1}
(see table 1 and below) in contrast to an anomalous first period
\te{\Delta_1\approx 1.8} and constant successive ones
\te{\Delta_2\approx\Delta_3 \approx\Delta_4 \approx 1.3} which were
quoted by Moshchalkov {\it et al.}~\cite{Moshchalkov}.

The field at which the ground state changes from \te{m=0} to \te{m=-1}
allows us to extract a lower critical field
\e{hcedisc}{H_{c1}^{\mbox{disc}} = 1.92376 \frac{\Phi_0}{\pi R^2}}
for a mesoscopic system with size \te{R} of order \te{\xi (0)}. Here \te{H_{c1}} is defined via the condition that for
\te{H<H_{c1}} the sample tries to screen out the applied flux, whereas
for \te{H>H_{c1}} the free energy is minimized by accepting one
flux quantum. It is interesting to compare this with Fetter's theory
of flux penetration in a superconducting disc \cite{Fetter}, which is
based on calculating the self energy of a vortex. In the limit where
the disc radius \te{R} is much smaller than the effective thin film
penetration depth \te{\lambda_{2d}= \lambda^2/d}, it turns out, that it
is energetically favourable for a vortex to enter if \te{H>H_{c1}} with
\cite{Fetter}
\e{hcefetter}{H_{c1} = \frac{\Phi_0}{\pi R^2} \ln \frac{R}{r_c} \qquad
\lambda_{2d} \gg R \gg r_c.} 
Here \te{r_c\approx \xi (0)} is the core radius, which is always
assumed to be much smaller than R. Obviously for samples whose size is
of the order of the coherence length \te{\xi(0)}, the expression
(\ref{hcefetter}) is no longer applicable. In this limit the
approximation that the order parameter is constant beyond \te{r_c}
becomes invalid. As found above the lower
critical field is then replaced by our result (\ref{hcedisc}), with a
crossover at about \te{R \approx 7 \xi (0) }. Here it is important
that for \te{ R \approx \xi (0)}, linearized GL-theory is sufficient
to calculate \te{H_{c1}}, because it is the sample boundary which
limits the magnitude of the order parameter instead of the quartic
term as usual. Finally consider a ring with inner radius \te{R_i \gg
r_c}. Then the lower critical field is simply determined by the
condition that half a flux quantum is applied, i.e.
\e{hcering}{H_{c1}^{\mbox{ring}} = \frac{1}{2} \frac{\Phi_0}{\pi R^2}.}
Indeed this follows from the quantization of the fluxoid \cite{Tinkham},
and is valid irrespective of the thickness of the ring.
Comparing (\ref{hcering}) with the result (\ref{hcedisc}) for a disc,
we find that \te{H_{c1}} in the latter case is almost four times larger. 
Qualitatively this is due to the additional condensation energy in the
center of the disc which is required for a vortex to enter.

As a second point let us discuss the behaviour at \te{\Phi \gg \Phi_0}
where many flux quanta have entered. In this limit the ground state
has angular momentum \te{|m|\gg 1}. The associated eigenfunction is
thus concentrated near the disc boundary. It is then obvious that our
upper critical field for \te{\Phi \gg \Phi_0} is in fact a surface critical field. If this is correct, it should
asymptotically approach the value obtained by Saint-James and de
Gennes \cite{SaintJamesDeGennes} for a surface with a radius of
curvature large compared to the coherence length. This can be verified by  considering the special values \te{\Phi_m, \,m=1,2,...} in Fig.~1,
where the tangent to \te{E_{0 |m|} (\Phi)} goes through the origin
(i.e.~we are considering successive approximations to the
envelope). These values are given in table 2 together with the
corresponding values of \te{\alpha_{max}}. 

\vspace{0.2cm}

\begin{tabular}{|r|r|r|r|}
\hline
& & & \\
$m$ & $\Phi / \Phi_0$ & $\alpha_{max}$ & $\frac{T_c - T_c(H)}{T_c}\, / \,\frac{\tilde{\Phi}}{\Phi_0}$  \\
\hline
1 & 2.44 & 0.28761 & 0.849  \\
2 & 3.92 & 0.26561 & 0.937  \\
3 & 5.28 & 0.25514 & 0.979  \\
4 & 6.56 & 0.24872 & 1.005  \\
5 & 7.82 & 0.24426 & 1.023  \\
6 & 9.09 & 0.24091 & 1.036  \\
7 & 10.28 & 0.23839 & 1.046  \\
8 & 11.46 & 0.23638 & 1.054  \\
9 & 12.68 & 0.23449 & 1.062  \\
10 & 13.86 & 0.23300 & 1.067 \\
100 & 110.23 & 0.21395 & 1.144  \\
200 & 214.72 & 0.21130 & 1.154 \\
1000 & 1033.76 & 0.20778 & 1.169  \\
10000 & 10109.33 & 0.20583 & 1.177 \\
\hline
\end{tabular}

TABLE 2

\vspace{0.2cm}

It is obvious that
\te{\alpha_{max}} converges to a limiting value 0.2058... . Using (\ref{dt})
 the associated transition temperature is then given by
\e{dtinf}{\frac{T_c - T_c (H)}{T_c}=1.177\frac{\tilde{\Phi}}{\Phi_0} .} 
As expected this is completely equivalent to the well known result 
\te{H=H_{c3} (T) = 1.695 H_{c2} (T)} for the surface critical field
\cite{SaintJamesDeGennes}. The coefficient \te{1 - 2 \alpha_{max}^m <
1} is in fact just the ratio between the ground state energy in the
disc and the energy \te{\hbar \omega_c/2}  of the lowest Landau
level in an infinite sample. Edge states centered near the sample
boundary have thus a \underline{lower} energy than bulk levels. Note
that this behaviour is just the opposite of the case with Dirichlet
boundary conditions (relevant e.g.~for edge states in the Quantum Hall
Effect) where edge states are \underline{above} the corresponding bulk
Landau levels \cite{Buisson}. Finally let us discuss the behaviour of
the periods \te{\Delta_j} of the ground state oscillations for large
flux \te{\Phi \gg \Phi_0}. Due to the factor \te{\rho^{|m|/2}} the order
parameter for increasing magnetic quantum number \te{|m|\gg 1} is more
and more concentrated near the sample boundary, but is practically
zero in the interior of the sample. The simply connected disc thus 
effectively behaves like a ring with a normal core of size
\te{R-c_1 l_H}, where \te{c_1} is a constant \cite{Buisson}. The
periodicity observed in \te{E_0 (H)} is then simply determined by the
condition that one additional flux quantum enters the area of the
normal core, i.e.
\e{flcond}{\Delta_j (\Phi\gg\Phi_0) = 1 \plu 2 c_1 \frac{l_H}{R}
\rightarrow 1.}
In fact this field dependance was observed in the experiments by Buisson et
al.~already for \te{\Phi/\Phi_0>5} \cite{Buisson}. In the asymptotic limit, which is 
however only reached for \te{\Phi/\Phi_0>10^3} (see table 2) the coefficient
\te{c_1} can be obtained analytically as \te{2c_1=\sqrt{0.59}\approx 0.76} 
\cite{SaintJames}.
\section{Discussion} 
Using linearized GL-theory we have calculated the nucleation field of a
small superconducting disc with a radius which is of the order of the
coherence length \te{\xi (0)}. The good agreement with the
experimentally observed $H-T$ phase boundary suggests that the
macroscopic GL-description remains valid in this regime which is not
obvious a priori. A surprising feature of our results is that
Aharonov-Bohm like oscillations are present even in a simply connected
sample. The physical origin of this effect is that already the
entrance of a single flux quantum effectively makes the sample a
multiply connected one. In the limit \te{\Phi \gg \Phi_0} the disc
behaves like a thin walled ring, leading to oscillations in \te{T_c (H)}
which are completely equivalent to the well known Little-Parks
experiment.
 It is interesting to note that these effects depend  crucially
on the Neumann boundary conditions. In fact the
equivalent eigenvalue spectrum with Dirichlet boundary conditions,
which was studied by Nakamura and Thomas \cite{Nakamura}, does not
exhibit any oscillations in the ground state energy \te{E_0 (\Phi)}.
It is an interesting future problem to investigate similar effects in
the fluctuation diamagnetism \cite{Schmidt} or extend the calculations
above to more complicated geometries. This would allow to study
eigenvalue spectra for systems with classical chaotic dynamics
\cite{Ullmo,Oppen} without the complications due to electron-electron
interactions which are unavoidable in non-superconducting mesoscopic systems.

\end{multicols}
\end{document}